\documentclass[a4paper]{jpconf}
\usepackage{graphicx}
\begin{document}
\title{Jet reconstruction and underlying event studies in p+p and d+Au collisions from STAR }

\author{Jana Bielcikova for the STAR Collaboration}

\address{Nuclear Physics Institute ASCR, Na Truhlarce 39/64, Prague 180 86, Czech Republic}

\ead{jana.bielcikova@ujf.cas.cz}

\begin{abstract}
Description of parton interaction with the hot and dense nuclear matter created in heavy-ion collisions at high energies is a complex task, which requires a detailed knowledge of jet production in p+p and d+Au collisions.  Measurements in these collision systems are therefore essential to disentangle initial state nuclear effects from cold nuclear
matter effects,  medium-induced $k_T$ broadening and jet quenching. To obtain complete description of the p+p (d+Au) collision it is also important to study particle production in the underlying event. The measured properties of underlying event can be used to tune the QCD based Monte-Carlo models. 

 In this paper some of the recent results on jet and underlying event properties in p+p and d+Au collisions at $\sqrt{s_{NN}}$~=~200 GeV measured by the STAR experiment are presented. In particular, the preliminary results on charged and strange particle fragmentation functions in p+p collisions are discussed and confronted with PYTHIA simulations. Next, the measurement of inclusive jet spectrum and di-jet correlations in d+Au collisions is presented and compared with the measurements in p+p collisions to estimate the size of cold nuclear matter effects. Finally, the study of underlying event properties in p+p collisions is shown and compared to PYTHIA simulation.
\end{abstract}

\section{Introduction}
Full jet reconstruction is considered as a promising tool for the
quantitative study of properties of the hot and dense medium produced in
heavy-ion collisions at RHIC energy.  The complexity of heavy-ion
collisions, where modifications of the fragmentation functions due to
interaction of partons with the hot and dense medium are expected,
requires inevitably a detailed understanding of jet production in  p+p and d+Au
collisions. 
Therefore measurements in these collision systems can be used as a baseline to disentangle initial state nuclear 
effects from cold nuclear matter effects,  medium-induced $k_T$ broadening and jet quenching.  In addition, 
it is also important to understand how the beam-beam remnants,
multiple parton interactions and initial- and final-state radiation
combine to produce particles in the underlying event (UE). The extracted
underlying event properties serve as in important input for tuning of the QCD
Monte-Carlo models such as PYTHIA.

The first results on inclusive jet spectra in p+p collisions at $\sqrt{s}$~=~200 GeV measured by the STAR experiment at RHIC in 2003-2004 
were published in~\cite{ppjetprl}. The measured inclusive jet cross-section is in a very good 
agreement with NLO pQCD calculations over seven orders of magnitude. 
During the data taking in 2006 and 2008, the STAR experiment collected large data
samples of p+p and d+Au collisions at $\sqrt{s_{NN}}$~=~200 GeV.
In this paper, the study of p+p collisions is extended by investigating fragmentation functions of charged and strange particles ($K^0_S$, $\Lambda$, $\bar{\Lambda}$) as well as underlying event properties. Further, the paper discusses recent preliminary results on inclusive jet spectra and di-jet correlations in d+Au collisions.

\section{Dataset and analysis}

The preliminary results presented in this paper are based on analysis of p+p collisions at $\sqrt{s}$~=200~GeV measured in 2006 and p+p and d+Au collisions at $\sqrt{s_{NN}}$~=200~GeV recorded in 2008 by the STAR experiment.  To increase yield of hard processes, two special triggers utilizing information from Barrel Electromagnetic Calorimeter (BEMC) were used in addition to minimum bias trigger. In 2006, a ``jet-patch'' trigger  defined as a $\Delta\eta\times\Delta\phi$~=~1$\times$1 patch of BEMC containing transverse energy $E_T>$~8~GeV was used. 
 In 2008, an online BEMC high tower (HT) trigger was used. To satisfy the HT trigger condition, at least one tower of BEMC must contain $E_T$ above a predefined threshold, which in case of the data presented here was 4.3~GeV.  In d+Au collisions, the Beam Beam Counter detector, located in the fragmentation region of Au nucleus, was used to select the 20\% highest multiplicity events. 

Jets were reconstructed using sequential recombination algorithms $k_t$ and anti-$k_t$ as well as seedless infrared safe cone algorithm SISCone, which are implemented in the FastJet package~\cite{fj1,fj2,fj3}. For the measurement of the neutral component of jets BEMC was used. The charged component of jets was measured by the Time Projection Chamber (TPC).  Both TPC and BEMC have a full azimuthal coverage and their combined pseudorapidity acceptance is $|\eta|<$~0.95. This limits the fiducial acceptance of jets in pseudorapidity to 1-$R$, where $R$ is the resolution parameter used in jet reconstruction. To avoid double counting of energy (electrons, minimum ionizing particles and possible hadronic showers in BEMC), the so called 100\% hadronic correction was applied: the momentum of a charged particle from TPC was subtracted off the neutral tower energy in BEMC, if the measured charged particle track pointed to a BEMC tower. In order to minimize biases due to dead areas and possible background in BEMC, the neutral energy fraction (NEF) of the total jet energy was required to be within 0.1~$<$~NEF~$<$~0.9. For charged particle tracks, an upper transverse momentum cut-off $p_T<$~15~GeV/$c$ was applied due to uncertainty in TPC tracking performance at large transverse momentum. Detailed investigation of this limitation is in progress.  


In parallel to experimental data analysis, the analysis of jets and underlying event properties was also performed on simulated data. The simulations are  based on PYTHIA 6.410, tuned to the CDF data at $\sqrt{s}$~=1.96~TeV (Tune A) and combined with a full STAR detector simulation based on  GEANT. The simulated data were then analyzed using the same analysis chain as that for the measured data. We distinguish several levels of the simulated data: jet reconstruction at Monte Carlo particle level (PyMC) and at detector level including GEANT simulation (PyGe). In addition, to determine influence of background and its fluctuations on jet production in d+Au collisions, PYTHIA events 
were embedded in 0-20\% highest multiplicity minimum bias d+Au events. This procedure was performed at detector level using reconstructed TPC tracks and signal in BEMC towers. This level of simulation is labeled as ``PyBg''.


\section{Fragmentation functions in p+p collisions}

\begin{figure}[t!] 
\begin{center}
\begin{tabular}{lr}
\includegraphics[height=5.7cm]{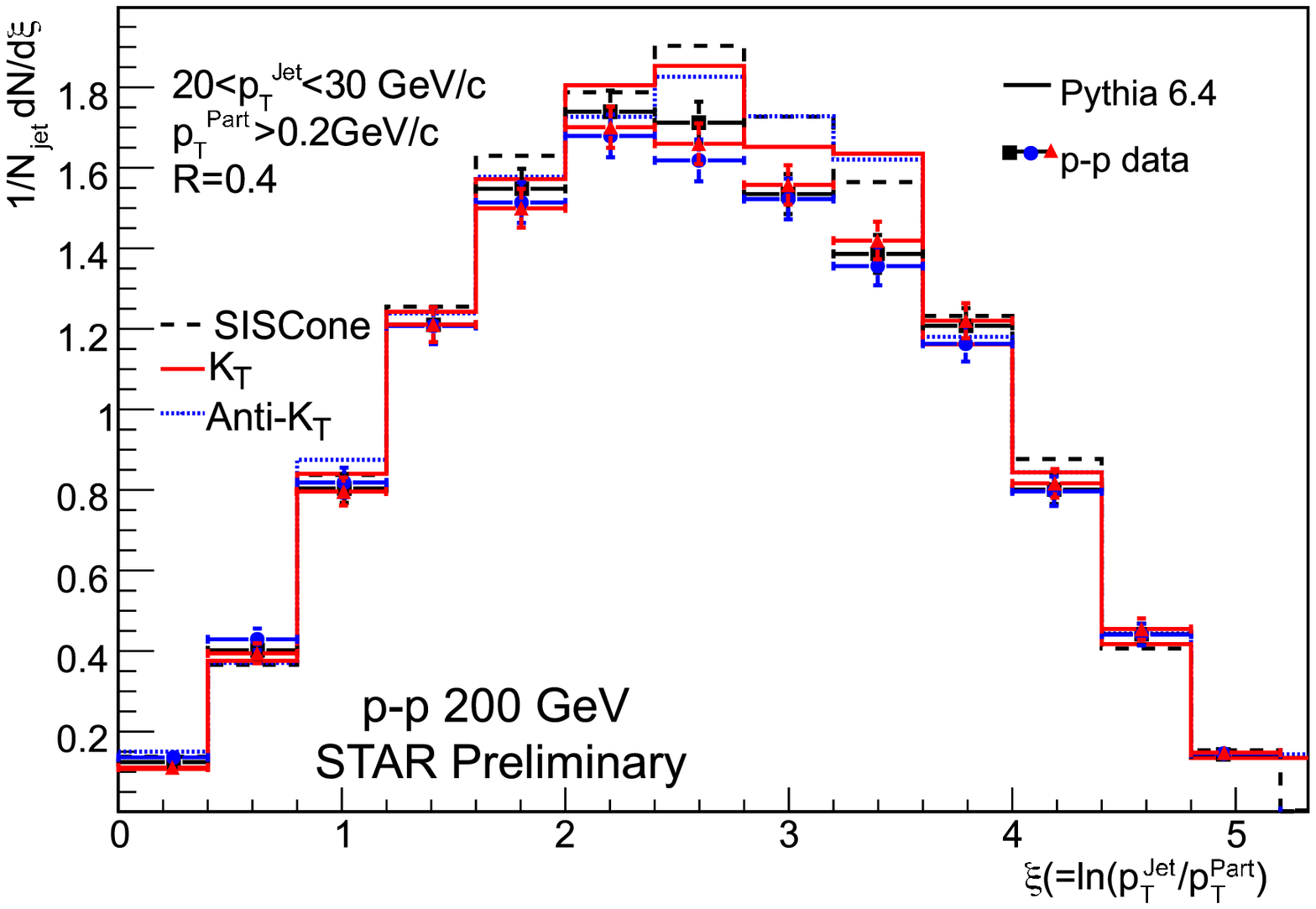}
&
\includegraphics[height=5.7cm]{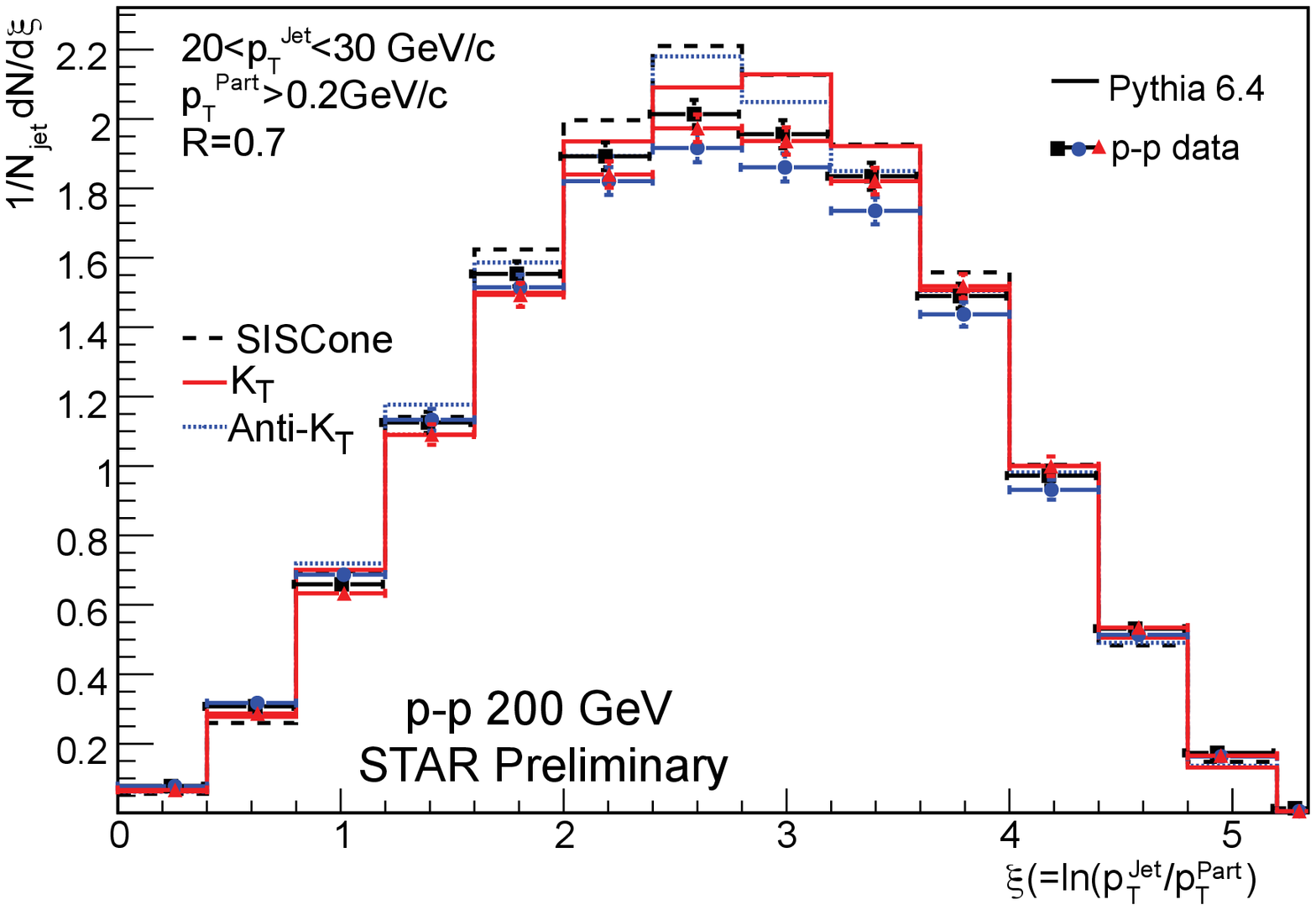}
\end{tabular}
\caption{Charged particle fragmentation function as a function of $\xi$ at detector level for jets with 20$<p_T<$~30~GeV/$c$ measured in p+p collisions at $\sqrt{s}$~=~200~GeV. The jets were reconstructed by three different jet algorithms (triangles: $k_{t}$,  circles: anti-$k_{t}$, squares: SISCone) with a resolution parameter $R$~=~0.4 (left) and $R$~=~0.7 (right)~\cite{CainesQM09,CainesDNP,CainesHP10}.  The data are compared to PYTHIA simulations (histogram).} 
\label{ff_charged_particles_r04}
\end{center}		
\end{figure}

Figure~\ref{ff_charged_particles_r04} shows the uncorrected fragmentation functions of charged particles for jets with transverse momentum  20~$<p_T<$~30~GeV/$c$ measured in p+p collisions at $\sqrt{s}$~=~200~GeV using the jet-patch triggered data~\cite{CainesQM09,CainesDNP,CainesHP10}.  As this trigger creates  a neutral energy fragmentation bias for the triggered jet, the fragmentation functions are shown only for the di-jet partner on away side. The fragmentation functions are plotted as a function of $\xi=\ln(p_{T}^{\mathrm{jet}}/p_{T}^{\mathrm{hadron}})$. The jets were reconstructed  by three different jet algorithms (k$_{t}$, anti-$k_{t}$ and SISCone) with a resolution parameter $R$~=~0.4 and $R$~=~0.7, respectively. All charged tracks measured by TPC and neutral towers from BEMC with $p_T (E_T)>$~0.2~GeV/$c$ were included in jet reconstruction. No significant dependence on jet algorithm is observed. The measured fragmentation functions are compared to PYTHIA simulations described above. For both resolution parameters studied there is a good agreement between the data and simulation. This suggests that the NLO contributions which are not included in PYTHIA are small at RHIC.

Further it is important to study in detail particle dependence of fragmentation functions as very little is known about identified fragmentation functions both experimentally and theoretically. While PYTHIA and NLO calculations agree reasonably well with the data on pion and proton production~\cite{AdamsPID}, theoretical description of strange particle production ($K$, $\Lambda$, $\Xi$) is a challenging task. The leading order calculations significantly underpredict the measured strange particle yields at intermediate and large transverse momenta~\cite{AbelevPID}. Higher-order corrections to leading-order pQCD cross sections, which are in general large, can be accounted for by introducing a $K$-factor, which renormalizes the leading-order pQCD cross section. It was found that a reasonable agreement of strange baryon particle production with PYTHIA at RHIC energy can be achieved by introducing $K$~=~3~\cite{AbelevPID}. However, such a large value of $K$ breaks the agreement of pion and proton data with PYTHIA.

\begin{figure}[t!] 
\begin{center}
\includegraphics[height=6.0cm]{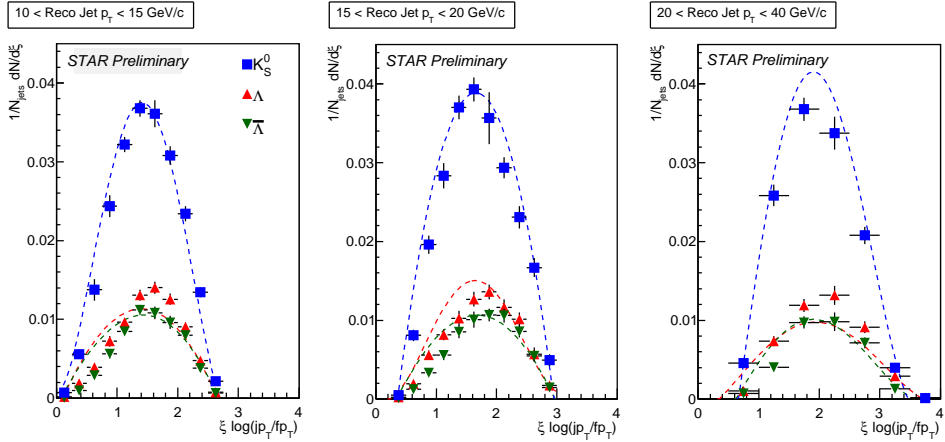}
\caption{Fragmentation functions of $K^{0}_{S}$ (squares), $\Lambda$ (triangles) and $\bar{\Lambda}$ (upside-down triangles)  as a function of the reconstructed jet $p_{T}$ in p+p collisions at $\sqrt{s}$~=~200~GeV~\cite{TimminsSQM}.  The curves show polynomial fits to PYTHIA 6.4 predictions. Both data and simulations are shown at detector level.}
\label{ff_strange_particles_r04}
\end{center}		
\end{figure}

The STAR experiment has excellent capability to reconstruct strange particles such as $K_0^S$ and $\Lambda$ via their typical V0 decay channels with a very good signal-to-noise ratio at $p_T>$~1~GeV/$c$.
Figure~\ref{ff_strange_particles_r04} shows the measured fragmentation functions  of $K^{0}_{S}$, $\Lambda$ and $\bar{\Lambda}$ particles in $p+p$ collisions at $\sqrt{s}$~=~200~GeV at detector level~\cite{TimminsSQM}. The fragmentation functions were calculated for three different jet $p_{T}$ ranges: 10-15, 15-20, 20-40~GeV/$c$. The remaining small residual background under the mass peaks is not yet corrected for.  The data are compared to polynomial fits to PYTHIA predictions. PYTHIA describes well the fragmentation functions of $K^0_S$, which is in agreement with measurements of $K^{\pm}$ fragmentation functions in $e^{+}+e^{-}$ collisions for similar jet energy range.  For $\Lambda$ and $\bar{\Lambda}$ fragmenation functions, PYTHIA predicts well the over-all yields, but it overestimates the yields at low $\xi$ and underestimates the yields at intermediate values of $\xi$. The integrated $\bar{\Lambda}/\Lambda$ and $\Lambda/K^0_S$ ratios fro $p_T>$~1~GeV/$c$ in the studied jet $p_T$ range are consistent with the values obtained from the minimum bias inclusive spectra when the same $p_T$ selection is applied~\cite{TimminsSQM}.  This observation indicates that hard processes significantly contribute to $K^0_S$, $\Lambda$, and $\bar{\Lambda}$ production at $p_T>$~1~GeV/$c$. Further studies are needed to verify this. Corrections of the data to particle level as well as removal of remaining combinatorial background under mass peaks are ongoing.

\section{Inclusive jet spectrum in d+Au collisions}
The inclusive transverse momentum spectrum of jets was measured in 10M 0-20\% highest multiplicity d+Au collisions at $\sqrt{s_{NN}}$~=~200~GeV. All charged tracks measured by  TPC and neutral towers from BEMC with $p_T>$~0.2~GeV/$c$ were included in jet reconstruction. The jets were reconstructed with the anti-$k_t$ jet finding algorithm with a resolution parameter $R$~=~0.4.  For subtraction of d+Au underlying event background, a method based on active jet areas~\cite{bgsub1,bgsub2} was applied on event-by-event basis. The inherent asymmetry of the d+Au collision system results in $\eta$ asymmetric underlying background. The $\eta$ dependence of underlying background was parametrized by a linear function and included in the background subtraction procedure.

  To correct for detector effects in d+Au collisions, the PYTHIA simulation of p+p collisions was used. In addition, a correction  accounting for the lower TPC reconstruction efficiency in d+Au relative to p+p collisions was applied to avoid bias due to different Jet Energy Scale (JES). The TPC reconstruction efficiency in d+Au was determined from a single charged pion embedding into measured minimum bias d+Au events on detector level. The higher tracking efficiency in the PYTHIA simulation was then artificially lowered to match the value in d+Au collisions. This procedure was applied prior to jet finding at PyGe and PyBg level of simulations. 

	To correct the jet spectrum to particle level, a bin-by-bin correction method was used. This correction procedure is based on the generalized efficiency determined as the ratio of PyMC to PyBg jet $p_T$ spectra, which is applied to the measured jet $p_T$ spectrum in d+Au collisions. The validity of this correction procedure relies on the fact that the shapes of the PyBg and the measured jet $p_T$ spectra are consistent. In~\cite{KapitanHP10} it was demonstrated that this condition is within statistical errors satisfied.

\begin{figure}[t!]
\begin{center}
\includegraphics[height=6.0cm]{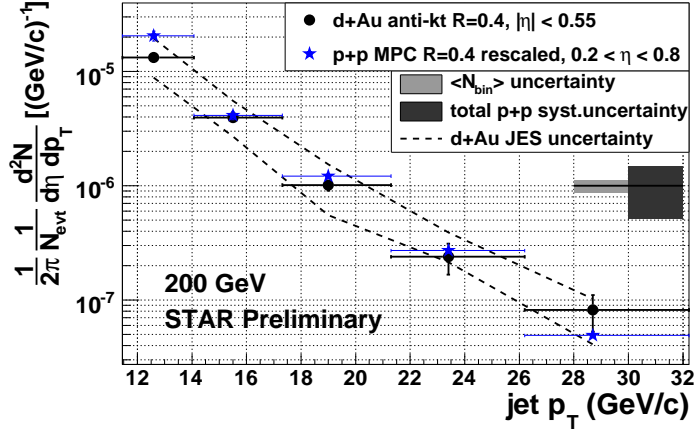}
\caption{\label{fig-dau-ptspec} Inclusive jet $p_T$ spectrum in 0-20\% highest multiplicity d+Au collisions at $\sqrt{s_{NN}}$~=200~GeV compared to $\langle N_{bin} \rangle$ scaled p+p spectrum from~\cite{ppjetprl}. The systematic errors (grey boxes and a dashed 
band) are explained in the legend.
}
\end{center}
\end{figure}

     The fully corrected inclusive jet $p_T$ spectrum in d+Au collisions at $\sqrt{s_{NN}}$~=~200~GeV is shown in Figure~\ref{fig-dau-ptspec}~\cite{KapitanHP10,KapitanHQ10}.  The systematic uncertainties are dominated by the JES uncertainty consisting of 10\% systematic uncertainty in TPC charged particle reconstruction efficiency and 5\% in BEMC calibration uncertainty, added in quadrature. Ongoing studies with a full jet embedding into measured d+Au events at raw detector level will enable to decrease the JES uncertainty. 

	The jet spectrum in d+Au collisions is compared with the jet $p_T$ spectrum in p+p collisions from 2003-2004 for which  a Mid Point Cone (MPC) jet algorithm with a cone radius $R$~=~0.4  was used~\cite{ppjetprl}. The effect of a slightly different $\eta$ acceptance on jet $p_T$ spectrum in case of p+p data was found to be less than 10\% in the covered $p_T$ range. For comparison with the jet spectrum in d+Au collisions, the jet spectrum in p+p collisions was scaled by an average number of binary collisions $\langle N_{bin} \rangle$~=~14.6$\pm$1.7 calculated from the Monte-Carlo Glauber model for 0-20\% highest multiplicity d+Au events. The jet spectra in both collision systems are consistent within current systematic errors dominated by the JES and MC Glauber model uncertainties.  A future comparison of the inclusive jet spectrum in d+Au collisions with that in p+p collisions measured during the same data taking period in 2008 is expected to further decrease the systematic uncertainties because they will be largely correlated. 
       


\section{Nuclear $k_T$ broadening}

Azimuthal correlations of di-jets can be used to quantify size of nuclear $k_T$ broadening in d+Au collisions relative to p+p collisions. The total transverse momentum of a di-jet ($k_{T,raw}$) can be determined from the di-jet opening angle $\Delta\phi$: $k_{T}$~=~$p_\mathrm{T}\sin(\Delta\phi)$. To increase the di-jet yield, the HT triggered dataset with $E_T>$~4.3~GeV was used for both p+p and d+Au data measured in 2008. Jets were reconstructed with the anti-$k_t$ algorithm using the resolution parameter $R$~=~0.5. In addition, a lower transverse momentum cut $p_T~>$~0.5~GeV/$c$ was applied to charged tracks and neutral towers to reduce background. For di-jet analysis the two most energetic jets in a given event ($p_\mathrm{T,1}>p_\mathrm{T,2}$) satisfying  10~$<p_\mathrm{T,2}<$~20~GeV/$c$ were selected. To obtain a clean di-jet sample a cut on di-jet  separation in azimuth $|\Delta\phi-\pi|<\pi/3$ was applied and 
measured $k_{T,raw}$ distributions were constructed as $k_{T,raw}$~=~$p_\mathrm{T,1}\sin(\Delta\phi)$.

\begin{figure}[t!]
\begin{center}
\includegraphics[height=6.0cm]{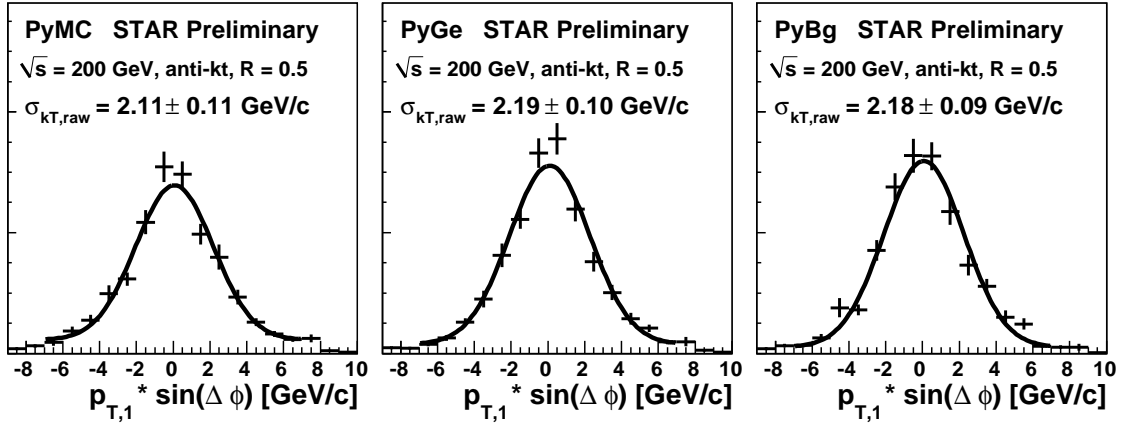}
\caption{
\label{fig:ktsimulation} Distributions of  $k_{T,raw}$~=~$p_\mathrm{T,1}\sin(\Delta\phi)$ for 10~$< p_\mathrm{T,2}<$~20~GeV/$c$ in MC simulated data PyMC (left), PyGe (middle) and PyBG (right).  Jets were reconstructed with the anti-$k_t$ algorithm with a resolution parameter $R$~=~0.5. The distributions were fit with a Gaussian fit function.}
\end{center}
\end{figure}
\begin{figure}[t!]
\begin{center}
\includegraphics[height=6.0cm]{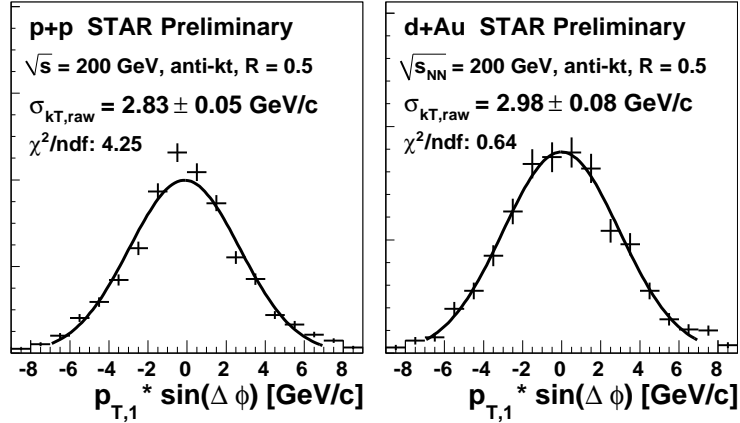}
\caption{\label{fig:ktdata} Measured distributions of  $k_{T,raw}$~=~$p_\mathrm{T,1}\sin(\Delta\phi)$ for 10~$< p_\mathrm{T,2}<$~20~GeV/$c$ in p+p (left) and 0-20\% highest multiplicity d+Au collisions (right) at $\sqrt{s_{NN}}$~=~200~GeV. Jets were reconstructed with the anti-$k_t$ algorithm with a resolution parameter $R$~=~0.5. The distributions were fit with a Gaussian fit function.}
\end{center}
\end{figure}
To quantify the size of detector and background effects on $k_{T,raw}$ distributions, a di-jet analysis was performed on PYTHIA simulated data which is demonstrated in Figure~\ref{fig:ktsimulation} together with a Gaussian fit of the simulated distributions~\cite{KapitanHP10,KapitanHQ10,KapitanQM09}. Comparison of PyMC, PyGe and PyBg distributions shows that the extracted Gaussian widths of $k_{T,raw}$ distributions are within statistical errors consistent. This suggests that detector and d+Au background effects in data can in first order be neglected.  It should be however noted, that in case of  PYTHIA simulations without background (PyMC and PyGe), the Gaussian fit does not describe the data well. 

 The $k_\mathrm{T,raw}$ distributions for data from p+p and d+Au collisions are shown in Figure~\ref{fig:ktdata}.
The values of Gaussian widths extracted from the fits are 
$\sigma_{kT,raw}^{p+p}$~=~2.8$\pm$0.1~(stat.)~GeV/$c$ in p+p collisions and
$\sigma_{kT,raw}^{d+Au}$~=~3.0$\pm$0.1~(stat.)~GeV/$c$ in d+Au collisions~\cite{KapitanHP10,KapitanHQ10,KapitanQM09}.
The widths in both collision systems are consistent within statistical errors indicating very small nuclear $k_T$ broadening. In case of p+p collisions, the Gaussian fit does not describe the shape of the $k_\mathrm{T,raw}$ distribution well similarly as observed for the PyMC and PyGe simulated data. The precise shape of the $k_\mathrm{T,raw}$ distribution is currently under investigation.
Varying the size of $|\Delta\phi - \pi|$ cut for back-to-back di-jet selection between 0.5 and 1.0, the systematic uncertainty on the extracted Gaussian widths of $k_\mathrm{T,raw}$ distribution was estimated to be 0.2~GeV/$c$. The effects of $p_\mathrm{T,2}$ range and jet finding algorithms were found to be negligible in both studied collision systems.

\section{Underlying event studies in p+p collisions}

\begin{figure}[t!]
\begin{center}
\includegraphics[height=6.0cm]{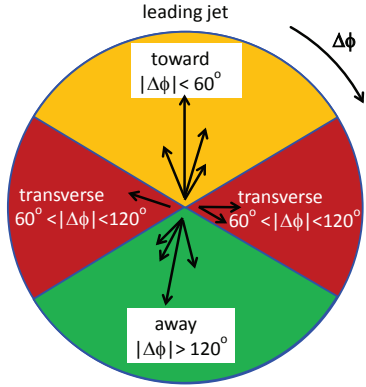}
\caption{Cartoon of the event areas (``toward'', ``away'', ``transverse'') in azimuthal angle difference relative to the leading jet ($\Delta\phi$).
\label{ue_cartoon}
}
\end{center}
\end{figure}

   A full description of particle production in p+p collisions  is a complex task which requires to incorporate both production of jets as well as production of particles in the so called underlying event within the same model. The underlying event (UE) is in general defined as everything but the hard scattering. It has contributions from soft and semi-hard multiple parton interactions (MPI), initial and final state radiation effects (ISR, FSR) and beam-beam remnants. As the study of UE presented below is focused on mid-rapidity, the contribution of beam-beam remnants to the UE properties can be neglected. 

   The UE study presented here follows the analysis technique developed by the CDF Collaboration at Tevatron~\cite{FieldCDF,CruzCDF}. After reconstructing jets in a given event, the event is split into four sections schematically depicted in Figure~\ref{ue_cartoon}. The sections are defined by their azimuthal angle difference  with respect to the leading jet axis ($\Delta\phi$). The ``toward'' region is the region 
within $|\Delta\phi|<$~60$^{\circ}$  around the leading jet axis and the ``away'' region is located at  $|\Delta\phi|>$~120$^{\circ}$. The remaining two regions labeled as ``transverse'' are placed at  60$^{\circ}<|\Delta\phi|<$~120$^{\circ}$ relative to the leading jet axis. The underlying event contributes to all of the mentioned regions, but to access its properties the two ``transverse'' regions are used as there the contamination from jets is assumed to be small. According to the charged particle multiplicity ($N_{ch}$) the two transverse regions are further labeled as ``TransMax'' and ``TransMin'', where $N_{ch}(\mathrm{TransMax})\geq N_{ch}(\mathrm{TransMin})$ in a given event.  This selection enhances contribution of initial and final state radiation components of the hard scattered parton in TransMax region relative to TransMin region.  The sensitivity to various physics processess in UE can be achieved 
by comparing properties of UE for a leading jet sample and a sample of events containing di-jets, which suppresses the probability of ISR and FSR effects. For this study  events with two reconstructed jets with $p_T^{away}/p_T^{lead}>$~0.7 and $|\Delta\phi|>$~150$^{\circ}$ were selected. 

\begin{figure}[t!]
\begin{center}
\begin{tabular}{lr}
\includegraphics[width=8cm]{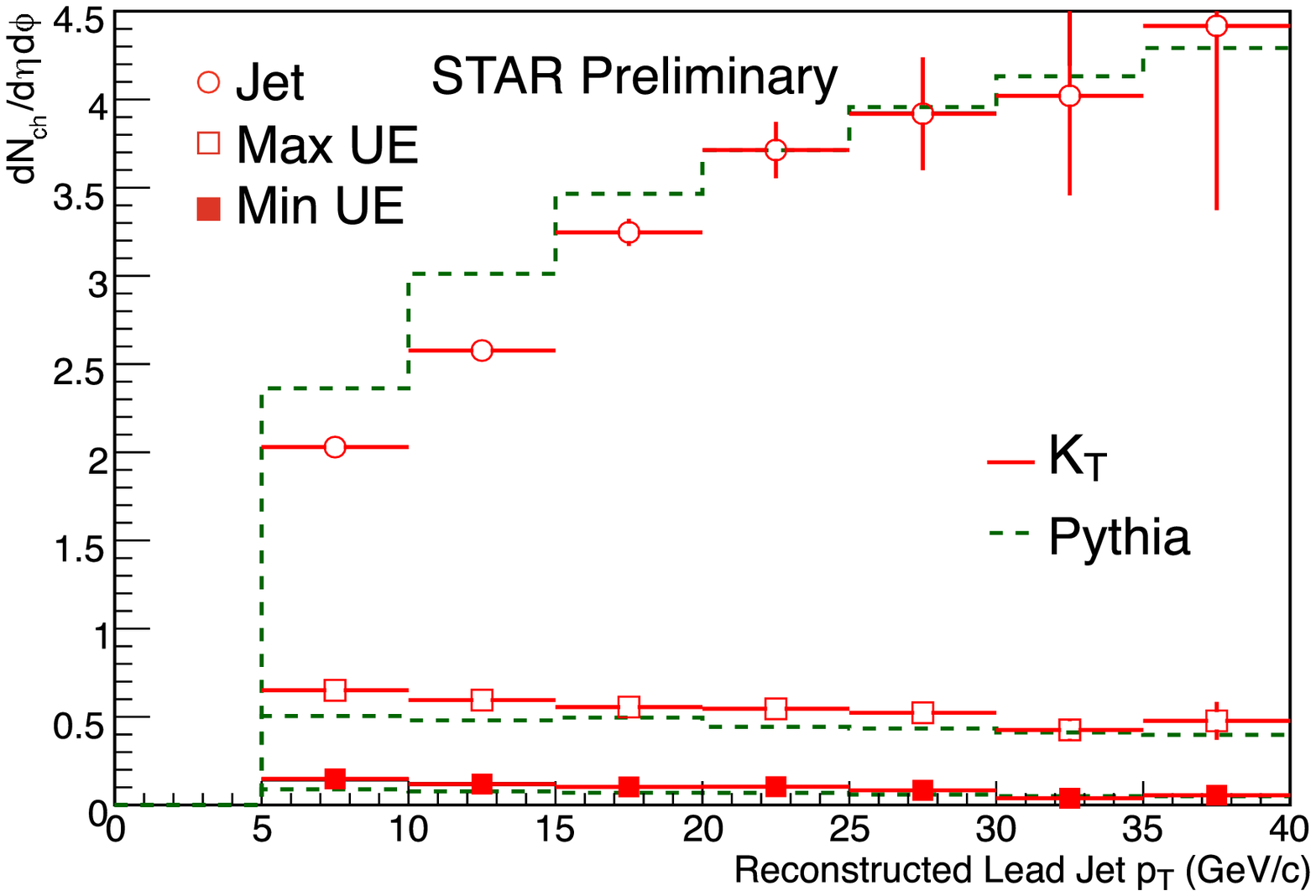}
&
\includegraphics[width=8cm]{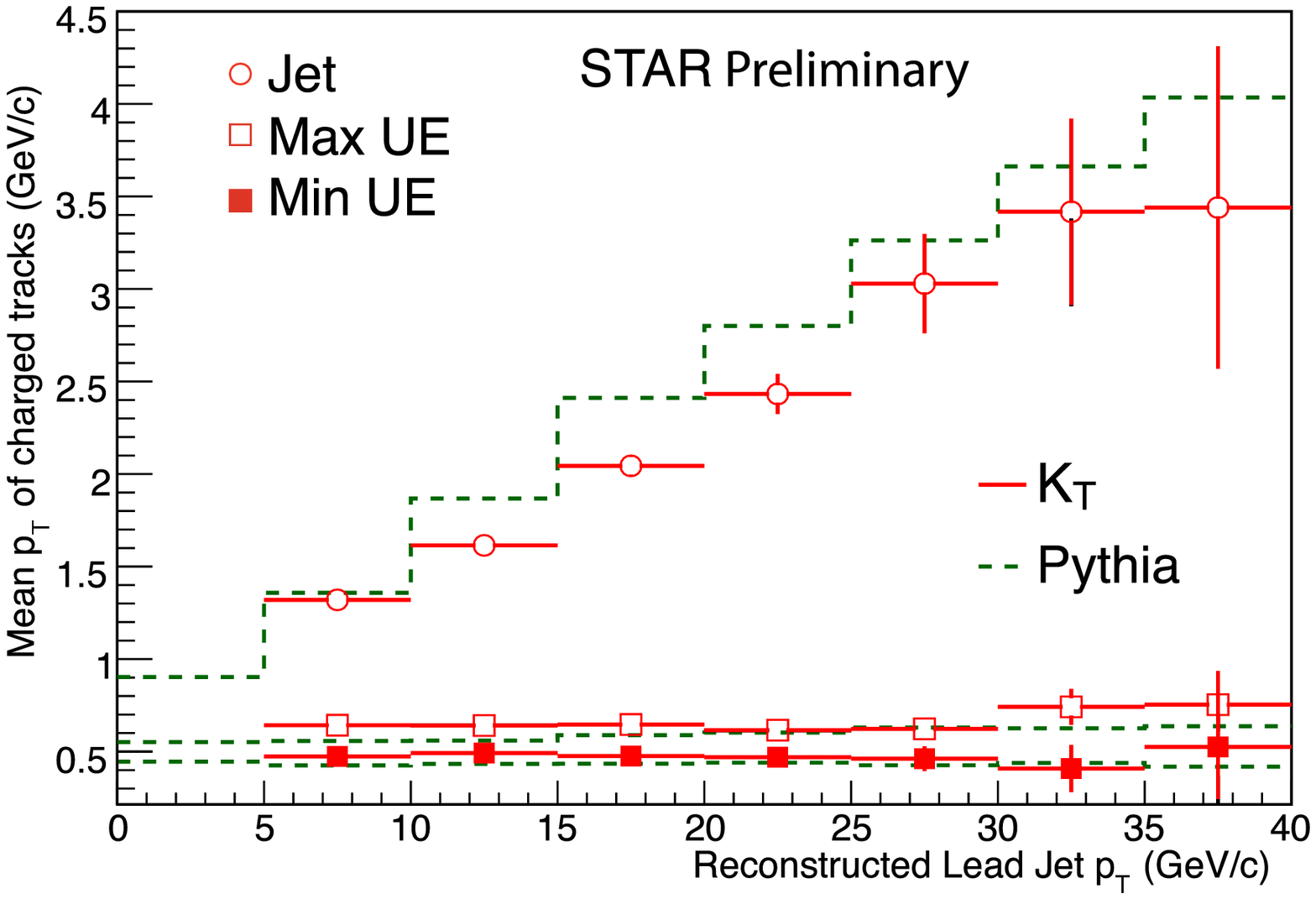}
\end{tabular}
\caption{The uncorrected charged particle density (left) and charged particle mean $p_T$ (right) in the ``di-jet" data set for the away side jet,  TransMin and TransMax regions as a function of reconstructed leading jet $p_T$. The jets were reconstructed with the $k_t$  algorithm with a resolution parameter $R$~=~0.7. The dashed histogram indicates the prediction from PYTHIA simulations (Tune A, $\epsilon$~=~0.25). The data and simulation are presented at detector level.}
\label{UE_nch_meanpt}

\end{center}
\end{figure}

  Figure~\ref{UE_nch_meanpt} shows the uncorrected charged particle density and mean transverse momentum in TransMin, TransMax and away side jet regions as a function of leading jet $p_T$ in  the di-jet sample. Jets were reconstructed with the $k_t$ algorithm with the resolution parameter $R$~=~0.7~\cite{CainesQM09,CainesDNP,CainesHP10}. No significant dependence on jet algorithm was observed. The charged particle density as well as  $\langle p_T \rangle$ are approximately independent of leading jet $p_T$, while the away-side jet charged particle density and $\langle p_T \rangle$  are rising with jet $p_T$, as expected. Compared to data measured at Tevatron ($\sqrt{s}$~=~1.96~TeV), the values of chaged particle density and $\langle p_T \rangle$ in the underlying event at RHIC energy are in general lower.

  The measured data in Figure~\ref{UE_nch_meanpt} are compared to PYTHIA simulation (Tune A) with MPI scaling factor $\epsilon$~=~0.25 based on recent measurements at Tevatron energies ($\sqrt{s}$~=~630, 1800 and 1960 GeV)~\cite{FieldAIP}. PYTHIA describes the main features of the data well, however a closer investigation reveals that PYTHIA slightly underestimates the charged particle multiplicity  and   $\langle p_T \rangle$ in TransMax and TransMin regions while overestimating these quantities within the jet.

\begin{figure}[t!]
\begin{center}
\includegraphics[height=6.0cm]{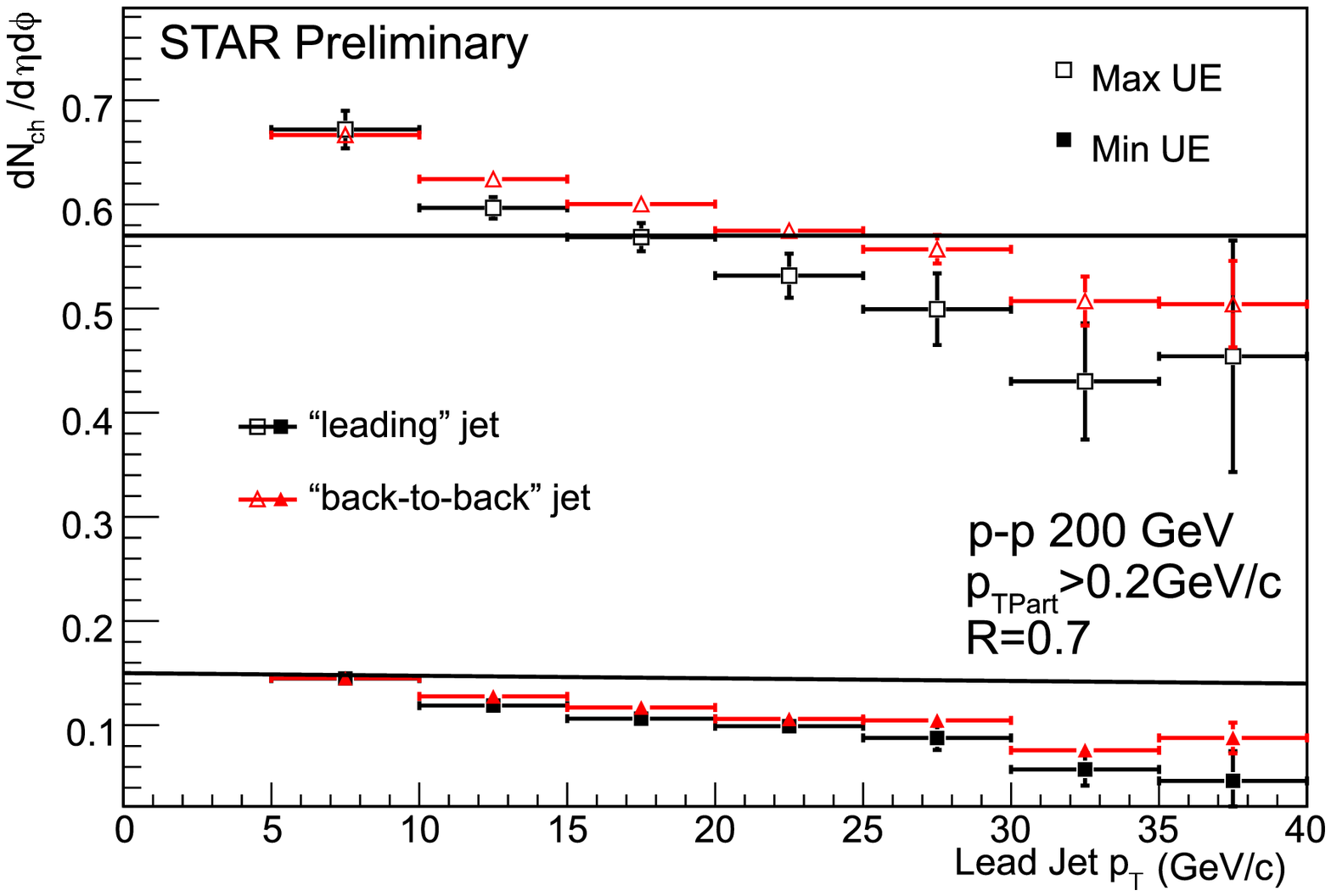}
\caption{The uncorrected charged particle density in the TransMin and TransMax regions as a function of reconstructed leading jet $p_T$ using SISCone algorithm with a resolution parameter $R$~=~0.7 shown separately for a leading jet and back-to-back jet samples. The data and simulation are presented at detector level. The solid line indicate expected charged particle density in case of simple Poisson distribution of charged particles in UE.}
\label{UE_nch_zoom_poisson}
\end{center}
\end{figure}

  In Figure~\ref{UE_nch_zoom_poisson} the measured charged particle density in UE is investigated closer by comparing the results for the transverse regions between the leading and di-jet samples~\cite{CainesQM09,CainesDNP,CainesHP10}. No significant difference between the jet samples is observed at RHIC energy which points to the fact that the hard scattered partons do not emit significant amount of large angle ISR and FSR. This observation is very different to that at Tevatron, where in $p+\bar{p}$ collisions at $\sqrt{s}$~=~1.96~TeV the ratio of charged particle density in leading/di-jet samples was found to be about 0.65~\cite{CruzCDF}. The data in Figure~\ref{UE_nch_zoom_poisson} are compared with an expected charged particle density in case the multiplicity in the UE would follow a simple Poisson distribution with an average value of 0.36. This simple simulation suggests that at RHIC energy, the splitting of the TransMax and TransMin values is mostly due to the sampling of the total charged particle multiplicity measured.

\section{Summary}

In summary, various properties of jets in p+p and d+Au collisions at $\sqrt{s_{NN}}$~=~200~GeV were discussed and compared to PYTHIA simulations. The measured fragmentation functions of charged particles and identified strange particles at detector level  show a good agreement with PYTHIA simulations for charged and $K^0_S$ particles, while for $\Lambda$ and $\bar{\Lambda}$ PYTHIA fails to describe details of the fragmentation pattern. 

The measured inclusive jet $p_T$ spectra and nuclear $k_T$ broadening in d+Au collisions from 2008 do not show any significant cold nuclear matter effects when compared to the measurements in $p+p$ collisions within current statistical and systematic uncertainties.  These uncertainties will be further decreased by comparing the data to a new jet cross section measurement in p+p collisions from 2008 and improvements in the tracking efficiency determination from jet embedding in real d+Au data.

First studies of underlying event properties in p+p collisions  show that its charged particle density and mean $p_T$ are largely independent of jet transverse momentum and the contributions from initial and final state radiation  are small. Studies of UE in d+Au collisions are underway and are expected to bring further insights into details of particle production mechanism in cold nuclear matter.

\vspace{0.5cm}

\section*{Acknowledgements}
\label{acknowledgement}

This work was supported in part by grants LC07048 and LA09013 of the Ministry of Education of the Czech Republic.


\end{document}